\documentclass[12pt]{article}

\usepackage{epsf,a4wide}

\renewcommand{\thefootnote}{\fnsymbol{footnote}}

\newcommand{\ub}{\bar u}
\newcommand{\xx}{\frac{m^2-\ub q^2}{u M^2}}

\begin{document}


\begin{titlepage}
\begin{flushright}
\begin{tabular}{l}
CERN--TH/01--257\\
DCPT/01/84\\
IPPP/01/42\\
ZU--TH 36/01
\end{tabular}
\end{flushright}
\vskip0.5cm
\begin{center}
  {\Large \bf 
Improved Analysis of $B\to\pi e \nu$ from\\[10pt] 
QCD Sum Rules on the Light-Cone}\\[1cm]
{\sc Patricia~Ball}${}^{1,}$\footnote{E-mail: Patricia.Ball@cern.ch, 
Patricia.Ball@durham.ac.uk}
and 
{\sc Roman Zwicky}${}^{2,}$\footnote{E-mail: zwicky@physik.unizh.ch, Roman.Zwicky@cern.ch}
\\[0.5cm]
\vspace*{0.1cm} ${}^1${\it CERN--TH, CH--1211 Geneva 23, Switzerland
\&\\ IPPP, University of Durham, Durham
  DH1~3LE, UK}\\[0.3cm]
\vspace*{0.1cm} ${}^2${\it Institut f\"ur theoretische Physik,
Universit\"at Z\"urich,
CH--8057 Z\"urich,
Switzerland \& CERN--TH, CH--1211 Geneva 23, Switzerland}\\[1.3cm]


  \vfill

{\large\bf Abstract\\[10pt]} \parbox[t]{\textwidth}{ 
We present a new calculation of the $B\to\pi$ form-factor $f_+$,
relevant for the measurement of $|V_{ub}|$ from semileptonic $B\to\pi$
transitions, from QCD sum rules on the light-cone. The new element is the
 calculation of radiative corrections to next-to-leading twist-3 accuracy.
  We find that these
contributions are factorizable at $O(\alpha_s)$, which lends
additional support to the method of QCD sum rules on the
light-cone. We obtain $f_+(0) = 0.26\pm 0.06 \pm 0.05$, 
where the first error accounts
for the uncertainty in the input-parameters and the second is a
guesstimate of the
systematic uncertainty induced by the approximations inherent in the
method. We also obtain a simple parametrization of the form-factor
which is valid in the entire kinematical range of semileptonic decays
and consistent with vector-meson dominance at large momentum-transfer.
}
  \vskip1cm 
{\em  submitted to JHEP}\\[1cm]
\end{center}
\end{titlepage}

\setcounter{footnote}{0}
\renewcommand{\thefootnote}{\arabic{footnote}}

\noindent {\large\bf 1.} The experimental programme of the dedicated
B-factories BaBar and Belle will contribute to unravel structure and
size of flavour- and CP-violation. The key-observable is the
unitarity-triangle of the CKM-matrix whose overdetermination will help
to answer the question whether there are additional sources of CP-violation
not present in the SM. Overdetermination means independent
measurements of sides and angles from different processes. One of the
sides is determined by the CKM-matrix element $|V_{ub}|$, whose
precise measurement is certainly a challenging task due to the
smallness of the corresponding branching ratios. The method of choice
is to measure it from semileptonic tree-level decays $b\to u e\nu$
where contamination by new-physics effects is expected to be 
small. The main complication in this measurement is, as a matter of course,
 QCD effects whose calculation from first principles is
highly challenging. Inclusive semileptonic decays are usually treated
in heavy quark expansion and become the more delicate the more accurately
experimentally necessary cuts are taken into account (which
necessitates the inclusion of other potentially large scales and calls
for threshold- and soft-gluon resummation, cf.~\cite{baetsch!}). The
alternative is to study exclusive decays where, owing to the
nonrenormalization of vector and axialvector currents, both perturbative and
nonperturbative QCD effects are neatly encoded in form-factors,
depending on only one variable, the momentum-transfer to the
leptons. The simplest such decay-process, involving only a single
form-factor\footnote{This statement is true only as long as
  lepton-masses are negligible, i.e.\ for semi-electronic and -muonic
  decays.}, is $B\to\pi \ell\nu_\ell$, which hence has received fair
attention in the literature. First experimental results are available
from CLEO \cite{CLEO}. The most precise calculation of the form-factor
will {\em sans doute} finally come from lattice-simulations; 
presently, however, the
main attention of lattice-practitioners appears to be directed not so much to 
obtaining 
phenomenologically relevant results, but rather to controlling lattice 
artifacts and approximations like discretization errors and the
explicit breaking of chiral symmetry with Wilson-fermions, 
particularly relevant for
calculating processes that involve pions; another problem is
how to simulate relativistic b-quarks, and how to access the region of
phase-space where the pion has large momentum. All these problems are
presently under intense debate, and we refer to Ref.~\cite{lattice}
for reviews and recent research papers.

Another, technically much simpler, but also 
less rigorous approach is provided by QCD sum rules on the
light-cone (LCSRs) \cite{BBK,CZ}. The key-idea is to
consider a correlation-function of the weak current and a current with
the quantum-numbers of the B-meson, sandwiched between the vacuum and
a pion. For large (negative) virtualities of these currents, the
correlation-function is, in coordinate-space, dominated by distances
close to the light-cone and can be discussed in the framework of
light-cone expansion. In contrast to the short-distance expansion
employed by conventional QCD sum rules \`a la SVZ \cite{SVZ}, where
nonperturbative effects are encoded in vacuum expectation values 
of local operators with
vacuum quantum numbers, the condensates, LCSRs
rely on the factorization of the underlying correlation function into
genuinely nonperturbative and universal hadron distribution amplitudes (DAs)
$\phi$ that are convoluted with process-dependent amplitudes $T_H$,
which are the analogues to the Wilson-coefficients in the
short-distance expansion and can be
calculated in perturbation theory, schematically
\begin{equation}\label{eq:1}
\mbox{correlation function~}\sim \sum_n T_H^{(n)}\otimes \phi^{(n)}.
\end{equation}
The sum runs over contributions with increasing twist, labelled by
$n$, which are suppressed by
increasing powers of, roughly speaking, the virtualities of the
involved currents. 
The same correlation function can, on the other hand, be written as a
dispersion-relation, in the virtuality of the current coupling to the
B-meson. Equating dispersion-representation and the
light-cone expansion, and separating the B-meson contribution from
that of higher one- and multi-particle states, one obtains a relation
for the form-factor describing $B\to\pi$. 

The particular strength of LCSRs lies in the
fact that they allow inclusion not only
of hard-gluon exchange contributions, which have been identified, in
the seminal papers that opened the study of hard exclusive processes
in the framework of perturbative QCD (pQCD)
\cite{pQCD}, as being dominant in light-meson form
factors, but that they also capture the so-called
Feynman-mechanism, where the quark created at the weak vertex carries
nearly all momentum of the meson in the final-state, while
all other quarks are soft. This mechanism is suppressed by two powers
of momentum-transfer in processes with light mesons; as shown in
\cite{CZ}, this suppression is absent in heavy-to-light
transitions\footnote{For very large quark masses, though, the
  Feynman-mechanism is suppressed by Sudakov-logarithms, which
  are, however, not expected to be effective at the b-quark
  mass.} and hence any reasonable application of pQCD to B-meson
decays should include this mechanism. LCSRs also avoid any reference to a
``light-cone wave-function of the B-meson'', which is a necessary
ingredient in all extensions of the original pQCD method
to heavy-meson decays
\cite{the_hard_guys,factorisation}, but whose exact definition
 appears to be problematic \cite{Korchemsky}.
A more detailed discussion of the
rationale of LCSRs and of the more
technical aspects of the method is beyond the scope of this
letter; more information can be found in the literature \cite{LCSRs:reviews}.

LCSRs are available for the $B\to\pi$ form
factor $f_+$ to
$O(\alpha_s)$ accuracy for the leading twist-2 contribution and at
tree-level for higher-twist (3 and 4) contributions
\cite{tree,LCSRs,rad}. 
In this letter we calculate the leading radiative corrections
to the twist-3 contributions. The motivation for this calculation is
twofold: first, it has been found in \cite{tree,LCSRs,rad} that the
tree-level twist-3 corrections are chirally enhanced and sizeable, 
and amount up to 30\% of
the final result for $f_+$, which indicates that 
radiative corrections may be
phenomenologically relevant. Second, the existence of the
factorization-formula  (\ref{eq:1}) is nontrivial beyond
tree-level and, to date, it is only for the twist-2 contribution that
factorization has been shown to hold also after inclusion of
radiative corrections. In this letter we show that (\ref{eq:1}), for a
certain approximation of the DAs $\phi^{(3)}$ (leading conformal spin), also
holds when $O(\alpha_s)$ corrections to the twist-3
contributions are included.

\bigskip

\noindent{\large\bf 2.} Let us now properly define the relevant
quantities. The form-factors $f_{+,0}$ are given by 
($q=p_B-p$)
\begin{equation}
\langle \pi(p) | \bar u \gamma_\mu b | B(p_B)\rangle  =  f_+(q^2) \left\{
(p_B+p)_\mu - \frac{m_B^2-m_\pi^2}{q^2} \, q_\mu \right\} +
\frac{m_B^2-m_\pi^2}{q^2} \, f_0(q^2)\, q_\mu;\label{eq:SL}
\end{equation}
in semileptonic decays the 
physical range in $q^2$ is $0\leq q^2\leq (m_B-m_\pi)^2$.
The starting point for the calculation of the form-factor
$f_+$ in (\ref{eq:SL})  is the
correlation function
\begin{eqnarray}\label{eq:CF}
i\int d^4y e^{iqy} \langle \pi(p)|T[\bar u\gamma_\mu b](y)
[m_b\bar b i\gamma_5 d](0)|0\rangle & = &
\Pi_+ 2p_\mu + \dots,
\end{eqnarray}
where the dots stand for structures not relevant for the calculation
of $f_+$. As mentioned before, for a certain configuration of
virtualities, namely $m_b^2-p_B^2\geq O(\Lambda_{\rm QCD}m_b)$ and $m_b^2-q^2\geq
O(\Lambda_{\rm QCD}m_b)$, the integral is dominated by light-like distances 
and accessible to an expansion around the light-cone:
\begin{equation}\label{eq:3}
\Pi_+ (q^2,p_B^2) = \sum_n \int_0^1 du\, \phi^{(n)}(u;\mu_{\rm IR}) 
T_H^{(n)}(u;q^2,p_B^2;\mu_{\rm IR}).
\end{equation}
As in (\ref{eq:1}), $n$ labels the twist of operators and 
$\mu_{\rm IR}$ is the (infrared) factorization-scale. The restriction
on $q^2$, $m_b^2-q^2\geq O(\Lambda_{\rm QCD}m_b)$, 
implies that $f_+$ is not accessible at all momentum-transfers; to be
be specific, we restrict ourselves to $0\leq q^2\leq 14\,$GeV$^2$.
As $\Pi_+$ is independent
of $\mu_{\rm IR}$, the above formula implies that the scale-dependence of
$T_H^{(n)}$ must be canceled by that of the DAs $\phi^{(n)}$. 

In (\ref{eq:3}) we have assumed that $\Pi_+$ can be described by
collinear factorization, i.e.\ that the only relevant degrees of
freedom are the longitudinal momentum fractions $u$ carried by the
partons in the $\pi$, and that
transverse momenta can be integrated over. Hard infrared (collinear) 
divergencies occurring in $T_H^{(n)}$ should be absorbable into the
DAs, as discussed in detail in
Ref.~\cite{Braaten}. Collinear factorization is trivial at tree-level,
where the b-quark mass acts effectively as regulator,
but can, in principle, be violated by radiative corrections, by
so-called ``soft'' divergent terms, which yield divergencies upon
integration over $u$. Such terms break for instance factorization in
non-leading twist in the treatment of nonleptonic B-decays \`a la BBNS
\cite{factorisation}; it is thus instructive to see what happens
in the simpler case of the correlation function (\ref{eq:CF}), where
the convolution involves only one DA instead of up
to three in $B\to\pi\pi$.
 To anticipate the result: we find that
factorization also works at one-loop level for twist-3 contributions
and that there are no soft divergencies.

There are two two-quark twist-3 DAs of the $\pi$, 
$\phi_p$ and $\phi_\sigma$, which are defined as
\begin{eqnarray}
\langle 0 | \bar u(x)[x,-x] i\gamma_5 d(-x) | \pi^-(p)\rangle & = &
\mu_\pi^2(\mu_{\rm IR})\, \int_0^1 du \, e^{i\xi px}\,
\phi_p(u,\mu_{\rm IR}),\label{eq:2.11}\\
\langle 0 | \bar u(x)[x,-x] \sigma_{\alpha\beta}\gamma_5 d(-x) |
\pi^-(p)\rangle & = & -\frac{i}{3}\, \mu_\pi^2(\mu_{\rm IR})
  (p_\alpha x_\beta-
p_\beta x_\alpha) \int_0^1 du \, e^{i\xi px}\,\phi_\sigma(u,\mu_{\rm IR})
\end{eqnarray}
with $\mu_\pi^2(\mu_{\rm IR}) = f_\pi m_\pi^2/(m_u+m_d)(\mu_{\rm IR})$ 
and $\xi = 2u-1$; $u$ is
the longitudinal momentum-fraction of the total momentum of the $\pi$ 
carried by the (d-)quark. 
The Wilson-line $$[x,-x]={\rm P}\exp[2ig\int_0^1 dtx_\mu A^\mu((2t-1)x)]$$ 
ensures gauge-invariance of the nonlocal matrix-elements. Exploiting
 conformal symmetry of massless QCD, which holds to leading-logarithmic
accuracy, one can perform a partial-wave expansion of the
DAs in terms of increasing conformal spin, which
amounts to an expansion in Gegenbauer-polynomials and 
fixes the functional dependence on $u$; the amplitudes of the
partial-waves are of nonperturbative origin and can be related to
local hadronic matrix-elements by virtue of the QCD equations of
motion as discussed in detail in \cite{wavefunctions,NLoperatoridentities}. 
It turns out, in particular, that the two DAs
$\phi_\sigma$ and $\phi_p$ are not independent, but mix with each
other and the twist-3 three-particle DA ${\cal T}$
that parametrizes the matrix-element
$\langle 0 | \bar u(x)[x,vx]\sigma_{\mu\nu}\gamma_5 g
G_{\alpha\beta}(vx)[vx,-x]
d(-x) | \pi^-(p)\rangle$. Thus, for consistency, when calculating
radiative corrections to $\Pi_+$ to twist-3 accuracy, one has to
include all three DAs,
$$
\Pi_+^{(3)} \sim \phi_p\otimes T_H^{(p)} + \phi_\sigma \otimes
T_H^{(\sigma)} + {\cal T}\otimes T_H^{({\cal T})},
$$
and it is only in the sum of these three terms that hard infrared divergencies and 
the scale-dependence are expected to cancel. The analysis of \cite{wavefunctions} has
 shown, however, that the two-quark DAs are very well
approximated by the lowest partial-wave, i.e.\ the one
 with smallest conformal spin, and that mixing with ${\cal
  T}$ sets in only at higher conformal spin. In calculating radiative
corrections to $\Pi_+^{(3)}$ we thus
restrict ourselves to leading conformal spin, i.e.\ the so-called
asymptotic DAs, and use \cite{wavefunctions}
\begin{equation}\label{eq:DAs}
\phi_\sigma(u) = 6 u (1-u),\quad \phi_p(u) = 1,\quad {\cal T} = 0.
\end{equation}

\bigskip

\begin{figure}
\centerline{\epsfxsize=0.9\textwidth\epsffile{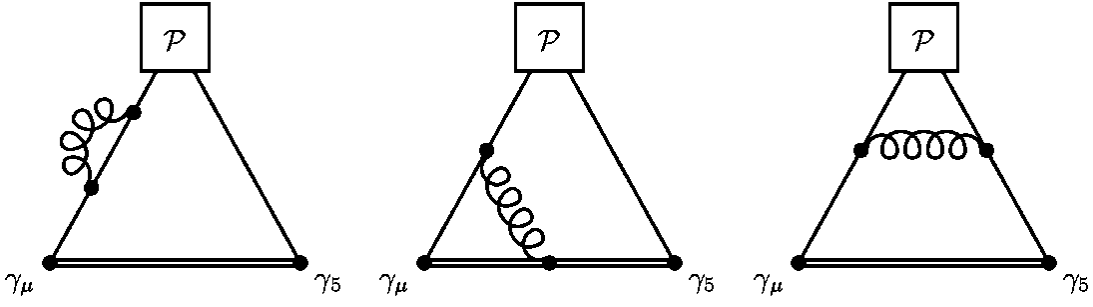}}
\caption[]{Some of the diagrams contributing to
$T_H^{(p,\sigma)}$ in one-loop order. The double line denotes the
b-quark propagator, the single lines denote the light u- and d-quark
propagators. $\gamma_\mu$ and $\gamma_5$ are the weak and the
B-vertex, respectively. There are two more self-energy and
one more vertex-correction diagrams.}\label{fig:1}
\end{figure}

\noindent{\large\bf 3.} Some of the diagrams contributing to
$T_H^{(p,\sigma)}$ to one-loop order are shown in Fig.~\ref{fig:1}.
The light quarks are massless and have momenta $up$ and $\ub p\equiv
(1-u) p$,
respectively. They are projected onto the desired DA
 by closing the trace with an appropriate projection
operator $\cal P$, which is just $-\mu_\pi^2 i\gamma_5/4$ for $\phi_p$ and
involves a derivative in $p$ for $\phi_\sigma$. The calculation
is performed in dimensional regularization for both ultraviolet and
infrared divergencies. Carefully distinguishing between the two types
of divergent terms, we find that the ultraviolet divergencies cancel
upon renormalization of the bare b-quark mass in the
tree-level expression, as they should. The infrared divergent terms,
on the other hand, do not cancel between $\mu_\pi^{2,{\rm
    bare}}\phi_{p(\sigma)}$ and $T_H^{(p(\sigma),{\rm bare})}$ separately, but
only in the sum of both contributions. The renormalized
$T_H^{(p,\sigma)}(u)$ are regular at
the endpoints, i.e.\ for $u\to 0$, $u\to 1$, which entails
the absence of soft divergent terms.

\begin{figure}
\centerline{\epsfxsize=0.55\textwidth\epsffile{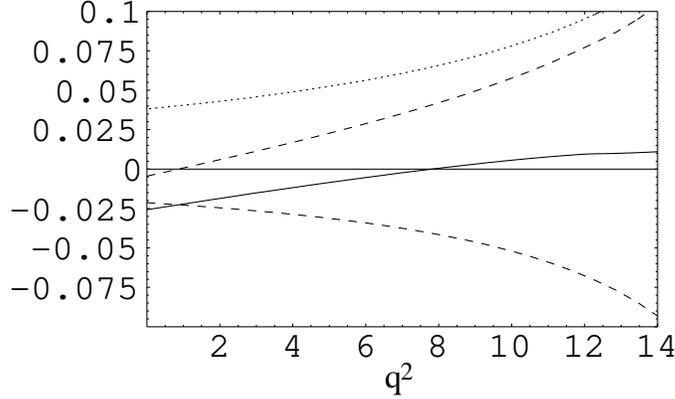}}
~\\[-1.3cm]
\caption[]{Radiative corrections to twist-2 and 3
  contributions, in units of $\alpha_s/(3\pi)$, 
for representative input-parameters, as functions  of
  $q^2$. Dashed lines:
  $T_H^{(p)}\otimes \phi_p$ and $T_H^{(\sigma)}\otimes \phi_\sigma$,
  respectively. Solid line: sum of dashed lines. Dotted line: 
 $T_H^{(2)}\otimes \phi^{(2)}$. The total twist-3 correction (solid
  line) is much smaller than the twist-2 correction (dotted
  line).}\label{fig:phi}
\end{figure}

As discussed below, the LCSR for $f_+$ involves the
continuum-subtracted Borel-transform $\hat{B}_{\rm sub} T_H$ 
of $T_H$. We calculate it by
splitting $T_H$ into two terms,
$
T_H = T_H^{\rm pole} + T_H^{\rm dis},
$
where $T_H^{\rm dis}$ can be written as a dispersion-relation in $p_B^2$
and its Borel-transform is obtained by applying Eq.~(\ref{eq:9}).
$T_H^{\rm pole}$ has a (single or double) pole in $s=m_b^2-u
p_B^2-\ub q^2\to 0$; the Borel-transforms of these terms are a bit
more involved, master-formulae are given in the appendix. The final
expressions for $\hat{B}_{\rm sub}T_H^{(p,\sigma)}$ are too bulky to
be presented here. A compact version of the tree-level expression can
be found in \cite{LCSRs}. In Fig.~\ref{fig:phi} we compare the
relative size of radiative corrections to twist-2 and 3
contributions. Whereas the absolute value of the two twist-3
corrections is, separately, of roughly the same size as that of the twist-2
correction, the only relevant quantity, the sum of both twist-3 corrections, is
much smaller than the twist-2 contribution, which indicates a good
convergence of the light-cone expansion also at $O(\alpha_s)$. 

The result of this calculation is the light-cone expansion of
$\Pi_+$, $\Pi_+^{\rm LC}$, and its continuum-subtracted Borel-transform, $\hat{B}_{\rm
  sub}\Pi_+^{\rm LC}$.

\bigskip

\noindent{\large\bf 4.} 
 Let us now derive the LCSR for $f_+$. The
correlation function $\Pi_+$, calculated for unphysical
$p_B^2$, can be written as dispersion-relation over its physical cut. Singling
out the contribution of the B-meson, one has
\begin{equation}\label{eq:corr}
\Pi_+ =  f_+(q^2) \, \frac{m_B^2f_B}{m_B^2-p_B^2}
+ \mbox{\rm higher poles and cuts},
\end{equation}
where $f_B$ is the leptonic decay constant of the B-meson,
$f_Bm_B^2=m_b\langle B| \bar b i\gamma_5 d|0\rangle$.
In the framework of LCSRs one does not use (\ref{eq:corr}) as it stands,
but performs a  Borel-transformation,
\begin{equation}\label{eq:9}
\hat{B}\,\frac{1}{t-p_B^2} = \frac{1}{M^2} \exp(-t/M^2),
\end{equation}
with the Borel-parameter $M^2$; this transformation enhances the
ground-state B-meson contribution to the dispersion-representation of $\Pi_+$
and suppresses contributions of higher twist to the light-cone expansion of
$\Pi_+$. The next step is to invoke quark-hadron
duality to approximate the contributions of hadrons other than the
ground-state B-meson by the imaginary part of the light-cone
expansion of $\Pi_+$, so that
\begin{eqnarray}
\hat{B}{\Pi_+^{\rm LC}} & = &
\frac{1}{M^2}\, m_B^2f_B \,f_+(q^2)\,e^{-m_B^2/M^2} +
\frac{1}{M^2}\, \frac{1}{\pi}\int_{s_0}^\infty \!\! dt \, {\rm
Im}{\Pi^{\rm LC}_+}(t) \, \exp(-t/M^2)\\
{\rm and}\quad \hat{B}_{\rm sub}\Pi_+^{\rm LC} & = & \frac{1}{M^2}\,
m_B^2f_B \,f_+(q^2)\,e^{-m_B^2/M^2}.\label{eq:SR}
\end{eqnarray}
Eq.~(\ref{eq:SR}) is the LCSR for $f_+$. 
$s_0$ is the so-called continuum
threshold, which separates the ground-state from the continuum
contribution. At tree-level, the continuum-subtraction in
(\ref{eq:SR}) introduces a lower limit of integration, $u\geq
(m_b^2-q^2)/(s_0-q^2)\equiv u_0$, in (\ref{eq:3}), which behaves as
$1-\Lambda_{\rm QCD}/m_b$ for
large $m_b$ and thus corresponds to the dynamical
configuration of the Feynman-mechanism, as it cuts off low momenta of
the u-quark created at the weak vertex. At $O(\alpha_s)$, there are
also contributions with no cut in the integration over $u$, which thus
correspond to hard-gluon exchange contributions. Numerically, these
terms turn out to be very small, $\sim O(1\%)$ of the total result for $f_+$.
As with standard QCD sum rules, the use of quark-hadron 
duality above $s_0$ and the
choice of $s_0$ itself introduce a certain model-dependence (or
systematic error) in the final result for the form-factor, which is
difficult to estimate. In this letter we opt for being rather conservative
and add a 20\% systematic error to the final result for $f_+$. Another
hadronic parameter showing up in (\ref{eq:SR}), which actually allows
one to fix the value of $s_0$, is $f_B$. $f_B$ can in principle be
measured from the decay $B\to \ell\bar\nu_\ell$, which, due to the expected
smallness of its branching ratio, BR$\,\sim O(10^{-6})$,
has, up to now, escaped experimental detection. $f_B$ is one of the
best-studied observables in lattice-simulations with heavy quarks; the
current world-average from unquenched calculations with two dynamical
quarks is $f_B = (200\pm 30)\,$MeV \cite{fBlatt}. It can also be
calculated from QCD sum rules: the most recent determinations
\cite{fBSR} include $O(\alpha_s^2)$ corrections and find $(206\pm 20)\,$MeV
and $(197\pm 23)\,$MeV, respectively. For consistency, we do not use
these results, but replace $f_B$ in (\ref{eq:SR}) by its QCD sum rule
to $O(\alpha_s)$ accuracy, including dependence on $s_0$ and $M^2$.
For the b-quark mass, we use an average over
recent determinations of the $\overline{\rm MS}$ mass,
$\overline{m}_{b,\overline{\rm MS}}(\overline{m}_b) = (4.22\pm
0.08)\,$GeV \cite{bquark,latmasses}, 
which corresponds to the one-loop pole-mass
$m_{b,{\rm 1L-pole}}=(4.60\pm 0.09)\,$GeV; this is different from the value
$m_b=4.8\,$GeV we used in our previous paper, the first reference in 
\cite{LCSRs}. With these values we find $f_B =(192\pm 22)\,$GeV (the
error only includes variation with $m_b$ and $M^2$, at optimized
$s_0$), in very good agreement with both lattice and QCD sum rules to
$O(\alpha_s^2)$ accuracy. For $m_b = (4.51,4.60,4.69)\,$GeV the optimized $s_0$ are
$(34.5,34.0,33.5)\,$GeV$^2$, and the relevant range in $M^2$ is 
$M^2\approx$(4.5--8)~GeV$^2$.

 The infrared
factorization-scale is set to $\mu_{\rm IR}^2 = m_B^2 - m_b^2$ \cite{bel}; the
dependence of $f_+$ on $\mu_{\rm IR}$ is very small, as all
numerically sizeable contributions are now available in
next-to-leading order in QCD, which ensures good cancellation of the
scale-dependence. 
For the $\pi$-DAs we use for the most part 
the same expressions as
in the first reference of \cite{LCSRs}, except for the new
$O(\alpha_s)$ corrections to $T_H^{(p,\sigma)}$, where we use the
DAs given in (\ref{eq:DAs}), and for the twist-2
DAs, where new analyses of the available experimental data on the
$\gamma^*\gamma\pi$ and the $\pi$ electromagnetic form-factor
indicate that $\phi^{(2)}$ is closer
to its asymptotic form than assumed previously and well
approximated by \cite{a2}
\begin{equation}\label{eq:xx}
\phi^{(2)}(u,\mu_{\rm IR}) = 6 u (1-u) \left\{ 1 +
a_2(\mu_{\rm IR}) \left( \frac{15}{2}\,(2u-1)^2
  -\frac{3}{2}\right)\right\}
\end{equation}
with $a_2(1\,{\rm GeV})=0.1\pm 0.1$.
The two-particle twist-3 DAs $\phi_p$ and $\phi_\sigma$ are
proportional to $\mu_\pi^2 = m_\pi^2 f_\pi/(m_u+m_d)$, which in the
chiral limit equals $- 2 \langle \bar q q\rangle/f_\pi$. Using the
``standard value'' of the quark-condensate, $\langle \bar q
q\rangle(1\,{\rm GeV}) = (-0.24\,{\rm GeV})^3$, one has
$\mu_\pi^2(\mu_{\rm IR}^2) = 0.25\,$GeV$^2$. With the central value for
the flavour-averaged light-quark mass, $\overline{m}_{ud}(2\,{\rm
  GeV}) = 4.5\,$MeV, from lattice \cite{latmasses}, one has
$\mu_\pi^2(\mu_{\rm IR}^2) = 0.30\,$GeV$^2$, with a quoted 
uncertainty
of 20\%. In our analysis we use the average $\mu_\pi^2(\mu_{\rm IR}^2) = 
(0.27\pm 0.07)\,$GeV$^2$.

\begin{figure}
\centerline{\epsfxsize=0.5\textwidth\epsffile{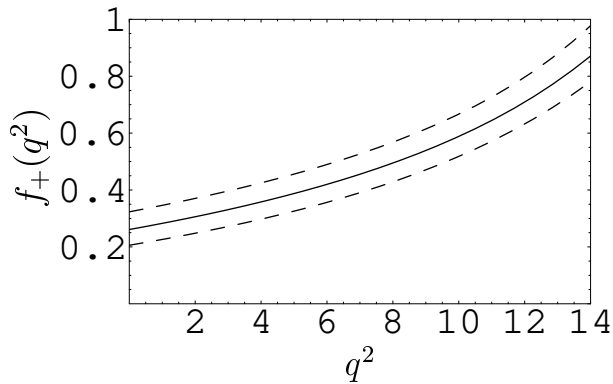}}
\caption[]{$f_+(q^2)$ as function of momentum-transfer $q^2$ from the LCSR
  (\protect{\ref{eq:SR}}). Solid line: LCSR for central values of
input-parameters and $M^2=6\,$GeV$^2$. Dashed lines: dependence of $f_+$ on
variation of input-parameters as specified in the text and
$5\,$GeV$^2\leq M^2 \leq 8\,$GeV$^2$.}\label{fig:2}
\end{figure}

With these parameters, we obtain the results shown in
Fig.~\ref{fig:2}. 
The form-factor can be accurately fitted by
\begin{equation}\label{eq:para}
f_+(q^2) = \frac{f_+(0)}{\displaystyle 1 - a\,(q^2/m_B^2) +
    b\,(q^2/m_B^2)^2}\,,
\end{equation}
with $f_+(0)$, $a$ and $b$ given in Tab.~1, for different values of $m_b$, $s_0$ and $M^2$.
The above parametrization reproduces the actual values calculated from the
LCSR, for $q^2\leq 14\,$GeV$^2$, to within 2\% accuracy.
At $q^2=0$ we find $f_+(0)=0.26\pm 0.06$, when all input-parameters
are varied within the ranges specified above. This has to be
compared to $0.31\pm 0.05$ from our previous analysis \cite{LCSRs}. At fixed
$M^2=6\,$GeV$^2$, and the central value for $m_b$, $m_b = 4.6\,$GeV,
we obtain, in this letter, $f_+(0) = 0.261$. Without the new radiative corrections,
the result becomes 0.294. Using in addition the parametrization of $\phi^{(2)}$
employed in \cite{LCSRs}, this becomes 0.321. And switching to
$m_b=4.8\,$GeV and $s_0 = 33.5\,$GeV$^2$, one obtains 0.308, i.e.\ the
central value for $f_+(0)$ quoted in \cite{LCSRs}. The tree-level
result for the central values of input-parameters is 0.247, i.e.\ the
total effect of radiative corrections is below 10\%.
\begin{table}
\addtolength{\arraycolsep}{3pt}
\renewcommand{\arraystretch}{1.2}
$$
\begin{array}{|l|l|l|l|}
\hline m_b [{\rm GeV}] &  4.69 &  4.60 & 4.51\\
s_0 [{\rm GeV}^2] & 33.5 & 34.0 & 34.5\\
a_2 (1\,{\rm GeV}) & 0 & 0.1 & 0.2\\
\mu_\pi^2(\mu_{\rm IR}^2)[{\rm GeV}^2] & 0.34 & 0.27 & 0.20\\
M^2 [{\rm GeV}^2] & 8 & 6 & 5\\\hline
f_+(0) & 0.206 & 0.261 & 0.323\\
a & 2.34 & 2.03 & 1.76\\
b & 1.77 & 1.29 & 0.87\\\hline
q_0^2 [{\rm GeV}^2] & 14.3 & 15.7 & 18.5\\
c_{\rm fit} & 0.384 & 0.439 & 0.523\\\hline
\displaystyle c_{\rm LCSR} = \frac{f_{B^*}g_{BB^*\pi}}{2m_{B^*}} &
0.396 & 0.414 & 0.430\\\hline
\end{array}
$$
\caption[]{Fit-parameters of Eqs.~(\protect{\ref{eq:para}}) and
(\protect{\ref{eq:Bstardom}}), including all variations of
input-parameters and $M^2$.}
\end{table}

It is instructive to compare the parametrization (\ref{eq:para}),
obtained at not too large
$q^2$, $q^2\leq 14 \,$GeV$^2$, with the vector-meson pole-dominance
approximation valid at large
$q^2\approx (m_B-m_\pi)^2 = 26.4\,$GeV$^2$: here $f_+$ is dominated
by the B$^*$-pole, located at $q^2 = m_{B^*}^2 = 28.4\,$GeV$^2$, and
can be expressed as
\begin{equation}\label{eq:Bstardom}
f_+(q^2) = \frac{c}{1-
  \frac{q^2}{m_{B^*}^2}}\,.
\end{equation}
The residue of the pole, $c$, can be related to physical couplings as
$c=f_{B^*} g_{BB^*\pi}/(2 m_{B^*})$, where $f_{B^*}$ is the
leptonic decay constant of the B$^*$ and $g_{BB^*\pi}$ is the coupling
of the B$^*$ to the B$\pi$-pair. $c$ can be calculated from LCSRs
itself: it is known up to twist-4 at tree-level; $O(\alpha_s)$
radiative corrections are known for the twist-2 contribution
\cite{bel,gBBpi}. 

\begin{figure}
\centerline{\epsfxsize=0.5\textwidth\epsffile{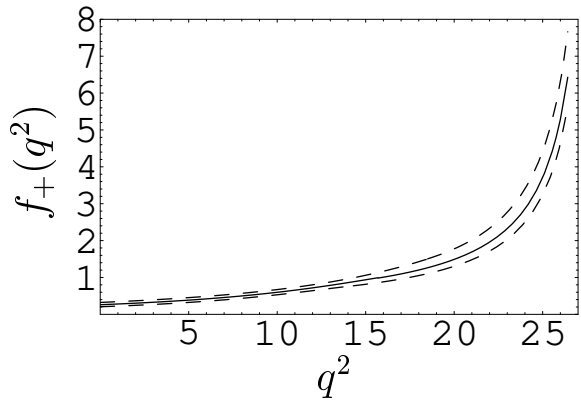}}
\caption[]{$f_+(q^2)$ as function of $q^2$ in the entire physical
range in $B\to\pi e \nu$, 
from Eqs.~(\protect{\ref{eq:para}}) and (\protect{\ref{eq:Bstardom}}),
including all variations of input-parameters and $M^2$. Solid line
obtained from 3rd column in Tab.~1, lower dashed curve from 2nd column
and upper dashed curve from 4th column.}\label{fig:4}
\end{figure}

We can now try to match the parametrizations (\ref{eq:para})
and (\ref{eq:Bstardom}). To this end, we treat $c$ as fit-parameter
and require that the transition
between both parametrizations be smooth, i.e.\ that at $q_0^2$ to
be fitted, the values of both $f_+(q^2_0)$ and its derivative are equal
for both parametrizations. The resulting values of $q_0^2$ and 
$c_{\rm fit}$ are
tabulated in Tab.~1, and the corresponding form-factors, obtained from
plotting (\ref{eq:para}) for $q^2\leq q_0^2$ and (\ref{eq:Bstardom})
for $q^2\geq q_0^2$, are shown in
Fig.~\ref{fig:4}.
The last row in Tab.~1 gives the values of $c_{\rm LCSR}$ 
obtained directly from the LCSR calculated in \cite{bel,gBBpi}. The agreement
between the direct and the fitted values is remarkably good, in view
of the fact that the LCSR for $c$ is less accurate than ours for
$f_+$, as it does not include $O(\alpha_s)$ corrections at
twist-3. Also the values of $q_0^2$ are well within expectation:
sufficiently below the pole on the one hand, but not too small on the
other hand.
Motivated by these results, we suggest a new parametrization of $f_+$
in terms of 5 parameters: Eq.~(\ref{eq:para}) for $q^2\leq q_0^2$ and
Eq.~(\ref{eq:Bstardom}) for $q^2\geq q_0^2$, with the set of
parameters given in Tab.~1 which comprises the full dependence of
$f_+(q^2)$ on input-parameters and the Borel-parameter.

\bigskip

\noindent {\large\bf 5.} In this letter we have investigated the
light-cone sum rule for the form-factor $f_+^{B\to\pi}$, including the
calculation of $O(\alpha_s)$ radiative corrections to the
next-to-leading twist-3 contribution. The calculation has demonstrated
the validity of the factorization formula (\ref{eq:3}) and the absence
of soft divergent terms to the considered accuracy, i.e.\ twist-3 to
$O(\alpha_s)$ with the DAs in the approximation of leading conformal
spin, Eq.~(\ref{eq:DAs}). As already found in \cite{rad}, the
Feynman-mechanism is the dominant contribution to $f_+$ and hard
perturbative corrections are numerically small.

In view of the systematic uncertainties inherent in LCSRs,
further refinement of the LCSR for $f_+^{B\to\pi}$ by including even 
higher twist contributions or more perturbative corrections 
is not likely to increase the overall accuracy. Improvement of the central
value of the result may, however, come from a reduced uncertainty of
input-parameters: whereas the dependence of $f_+(q^2)$ on $m_b$ is
rather small, a reduction of uncertainty in the
non-leading conformal spin contributions to $\phi^{(2)}$ would be
useful. As for the approximation of leading conformal spin DAs in
twist-3, Eq.~(\ref{eq:DAs}), at tree-level it yields about 90\% of
the actual result, which strengthens confidence that this
approximation works similarly well also at $O(\alpha_s)$. 
The change in central values of $f_+(q^2)$ as
compared to our previous results, Refs.~\cite{LCSRs,rad}, is partially
due to the $O(\alpha_s)$ corrections to twist-3 we have calculated in
this letter and partially due to updated input-parameters.

The LCSR is valid only for large
energies of the $\pi$, i.e.\ not too large values of $q^2$. In this letter we
have fixed, somewhat arbitrarily, the maximum allowed value of
$q^2$ at $q_{\rm max}^2$ at 14$\,$GeV$^2$ and have parametrized the
form-factor by Eq.~(\ref{eq:para}). On the other hand, for
$q^2$ close to the kinematical maximum allowed in semileptonic decays,
$q^2 = 26.4\,$GeV$^2$,
the form-factor is dominated by the close-by B$^*$-pole and can be
parametrized by Eq.~(\ref{eq:Bstardom}). The
residue of that pole can also be calculated from LCSRs. We have tried to match
both parametrizations, Eqs.~(\ref{eq:para}) and (\ref{eq:Bstardom}), by requiring
smoothness at the matching-point $q^2 = q_0^2$, which is
a parameter of the fit itself. The resulting
values of $q_0^2$ are  within expectations, and the fitted values
of the residue agree well with the direct calculation from LCSRs, as
demonstrated in Tab.~1. This result lends additional confidence to our
final parametrization of the form-factor, i.e.\ the combination of
Eqs.~(\ref{eq:para}) and (\ref{eq:Bstardom}), with a total of 5
parameters,
which is valid in the
complete range of kinematically allowed $q^2$ in $B\to\pi e \nu$, $0\leq q^2\leq
(m_B-m_\pi)^2 = 26.4\,$GeV$^2$.

\section*{Appendix}

\setcounter{equation}{0}
\renewcommand{\theequation}{A.\arabic{equation}}

In this appendix we collect formulas for non-standard Borel-transformations.
Generally, the Borel-transform $\hat{B}f(P^2)$ of a function
$f(P^2)$ of the Euclidean momentum $P$ is defined as 
$$
\hat{B}f(P^2) =
\lim_{\begin{array}{l}\scriptstyle
P^2\to\infty,\: N\to\infty\\[-5pt]\scriptstyle
P^2/N=M^2 {\rm ~fixed~}\end{array}}\,\frac{1}{N!}\, (-P^2)^{N+1}
\,\frac{d^{N+1}}{(dP^2)^{N+1}}\, f(P^2).
$$
By $\hat{B}_{\rm sub}f(P^2)$ we denote the Borel-transform including
continuum-subtraction above the threshold $s_0$, i.e.\ if $f$ has the
dispersion-representation (here $p$ is Minkowskian)
$$
f(p^2) = \int\limits_{m^2}^\infty dt\,\frac{\rho(t)}{t-p^2},
$$ 
we define
\begin{equation}\label{eq:app}
\hat{B}_{\rm sub}f(p^2) = \frac{1}{M^2}\,\int\limits_{m^2}^{s_0} dt\,\rho(t)\,
e^{-t/M^2}.
\end{equation}
We need in particular the following transforms
($s=m^2-u p^2-\ub q^2$):
$$
\hat{B}\,\frac{1}{(t-p^2)^\alpha\, s^\beta} =
\frac{1}{\Gamma(\alpha+\beta)}
  \,\frac{1}{u^\beta(M^2)^{\alpha+\beta}} \,e^{-t/M^2}\,
  {}_1F_1\left(\beta,\alpha+\beta, -\frac{m^2-ut-\ub q^2}{uM^2}\right),
$$
from which the Borel-transforms of expressions with additional
logarithms are obtained as derivatives, e.g.\ $\hat{B}s^{-\beta}\ln s =
-\frac{d}{d\beta}\, \hat{B}\,s^{-\beta}$. 
Including continuum subtraction, we find
$$
\hat{B}_{\rm sub}\,\frac{1}{s^\beta} =
\frac{e^{-\xx}}{(uM^2)^{\beta}}\, \frac{1}{\Gamma(\beta)} \left( 1 -
  \frac{\Gamma(1-\beta, \frac{u s_0+\ub q^2-m^2}{u
      M^2})}{\Gamma(1-\beta)}\right) \Theta(u-u_0)
$$
with $u_0 = (m^2-q^2)/(s_0-q^2)$. For integer $\beta$, the second
terms becomes a sum over $\delta(u-u_0)$ and its derivatives.
We also give the spectral function $\rho$ of the general expression 
$(m^2-p^2)^{-\alpha}s^{-\beta}$, from which the Borel-transform
$\hat{B}_{\rm sub}$
including continuum subtraction can be obtained using (\ref{eq:app}):
\begin{eqnarray*}
\rho(t) &=&
\Theta(\tilde{m}^2-t)\Theta(t-m^2)\,\frac{\sin\alpha\pi}{\pi}\,
\frac{1}{(t-m^2)^\alpha\, u^\beta(\tilde{m}^2-t)^\beta}\\ &&{}+
\Theta(t-\tilde{m}^2)\, \frac{\sin(\alpha+\beta)\pi}{\pi}\,
  \frac{1}{(t-m^2)^\alpha\, u^\beta(t-\tilde{m}^2)^\beta}
\end{eqnarray*}
with $\tilde{m}^2 = (m^2-\ub q^2)/u$.

\section*{Acknowledgements}

P.B.\ is supported by DFG (Deutsche Forschungsgemeinschaft) through a Heisenberg-fellowship. 
The work of R.Z.\ is supported by the Schwei\-ze\-ri\-scher Nationalfond.

\end{document}